\documentclass[prb,twocolumn,10pt,aps]{revtex4-1}
\usepackage[dvips]{graphics,graphicx}
\usepackage{amsmath}

\usepackage[russian,english]{babel}
\usepackage{amssymb}
\usepackage{epstopdf}
\usepackage[usenames, dvipsnames]{color}
\usepackage{graphicx}
\usepackage[utf8]{inputenc}

\usepackage{amsfonts}
\usepackage{eufrak}
\usepackage{mathrsfs}
\usepackage{bm}

\begin{document}

\title{Enhanced light absorption in Tamm metasurface with a bound state in the continuum}
\author{Rashid G. Bikbaev$^{1,2,*}$}
\author{Dmitrii N. Maksimov$^{1,2,*}$}
\author{Pavel S. Pankin$^{1,2}$}
\author{Ming-Jyun Ye$^{3}$}
\author{Kuo-Ping Chen$^{3,4}$}
\author{Ivan V. Timofeev$^{1,2}$}

\affiliation{$^1$Kirensky Institute of Physics, Federal Research Center KSC SB
RAS, 660036, Krasnoyarsk, Russia}
\affiliation{$^2$Siberian Federal University, Krasnoyarsk 660041, Russia}
\affiliation{$^3$College of Photonics, National Yang Ming Chiao Tung University, Tainan 711, Taiwan}
\affiliation{$^4$Institute of Photonics Technologies, National Tsing Hua University, Hsinchu 300, Taiwan}

\date{\today}
\begin{abstract}
We consider light absorption in a germanium grating placed on top of photonic-crystalline substrate. Such a system supports an optical Tamm state decoupled from the continuous spectrum with its frequency within the photonic band gap. We have demonstrated that application of the Tamm state makes in possible to engineer extremely narrow absorber which provides
a $100\%$ absorption in a semiconductor grating in the critical coupling regime. The proposed design
may be used at both normal and oblique incidence at the telecom wavelength. 
\end{abstract}
\maketitle

Localized modes in periodic structures are of great interest due to record high Q-factors~\cite{Akahane2003,Song2005,Asano2017} as well as the opportunity to design tunable devices by using liquid crystals~\cite{Hsiao2019}, nonlinear media~\cite{Soljai2004} or resonant materials~\cite{Moiseev2019}. 
Along with studies on volume electromagnetic waves localized in structural defects, attention is paid to surface waves localized at the boundary with negative dielectric permittivity media. In this case, the light is trapped at the boundary between plasmonic and dielectric structures. These localized states are called Tamm plasmon polaritons~\cite{Kaliteevski2007} (TPPs). The TPPs are applied for engineering optical devices, such as photoelectrochemical cells~\cite{Pyatnov2022}, sensors~\cite{Fecteau2021}, lasers~\cite{Xu2021}, beam steerers~\cite{Bikbaev2022Materilas} and solar cells~\cite{Bikbaev2020OSC}. Coupling of TPPs with other types of localized modes, for example, with a surface plasmon-polariton~\cite{Afinogenov2013b}, leads to hybrid modes which are widely applied in optical sensors\cite{Das2015,BuzavaiteVerteliene2020}. Tailoring the parameters of TPP supporting structures makes it possible to set-up the critical coupling at which all radiation incident on the structure is absorbed at the TPP wavelength. This resonant absorption mechanism is used in absorbers~\cite{Kim2022,Bikbaev2019Epsilon-Near-ZeroPolariton} and photodetectors~\cite{Wang2020,Zhang2017}. 
Such surface-localized status can be excited at the boundary between a photonic crystal and plasmonic~\cite{Buchnev2020,Buchnev2022} or dielectric metasurfaces~\cite{Bikbaev2021OptExp}. In the latter case, such localized states are called optical Tamm states.  

\begin{figure}[t]
\centering
\includegraphics[width=80mm]{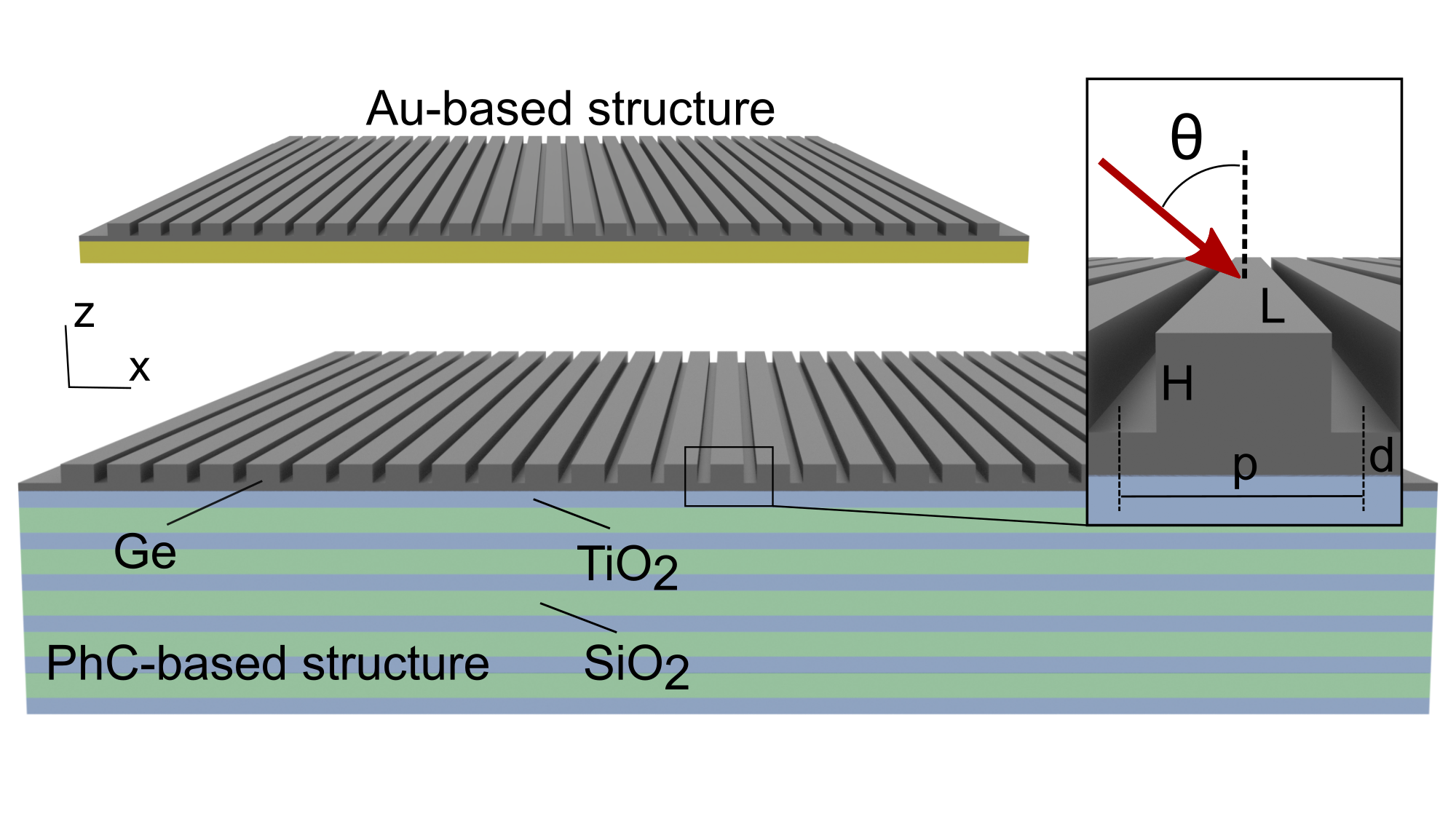}
\caption{Schematic representation of the Au and PhC-based structures. $H$ and $L$ are height and width of nanostripes. $p$ is period of nanostripes along $x$-axis and $d$ is Ge substrate thickness. The PhC consists of alternating layers of silicon dioxide and titanium dioxide with refractive indices 1.44 and 2.45, respectively. The thicknesses of layers are 268.3~nm and 156.8~nm. The thickness of the Au substrate is 200~nm. 
}
\label{fig.Fig1}
\end{figure}

Lately, we have seen a surge of interest to photonic non-radiation states, i.e. optical bound states
in the continuum (BICs)~\cite{Hsu16, Koshelev19, sadreev2021interference} which have become an important
instrument for engineering optical devices with enhanced light-matter interaction. In the presence of material absorption the BIC is shown to acquire finite-life, albeit remain localized and decoupled from the outgoing channels~\cite{Hu20}.
Quasi-BIC in lossy periodic structures is found to be instrumental for enhancement of light absorption~\cite{Saadabad21, Tan22}
in the critical coupling regime even in low loss dielectrics. Therefor, BIC concept opens novel opportunities for highly
efficient light absorbers \cite{zhang2015ultrasensitive, wang2020controlling, sang2021highly, xiao2021engineering, cai2022enhancing}.

\begin{figure*}[ht]
\centering\includegraphics[width=175mm]{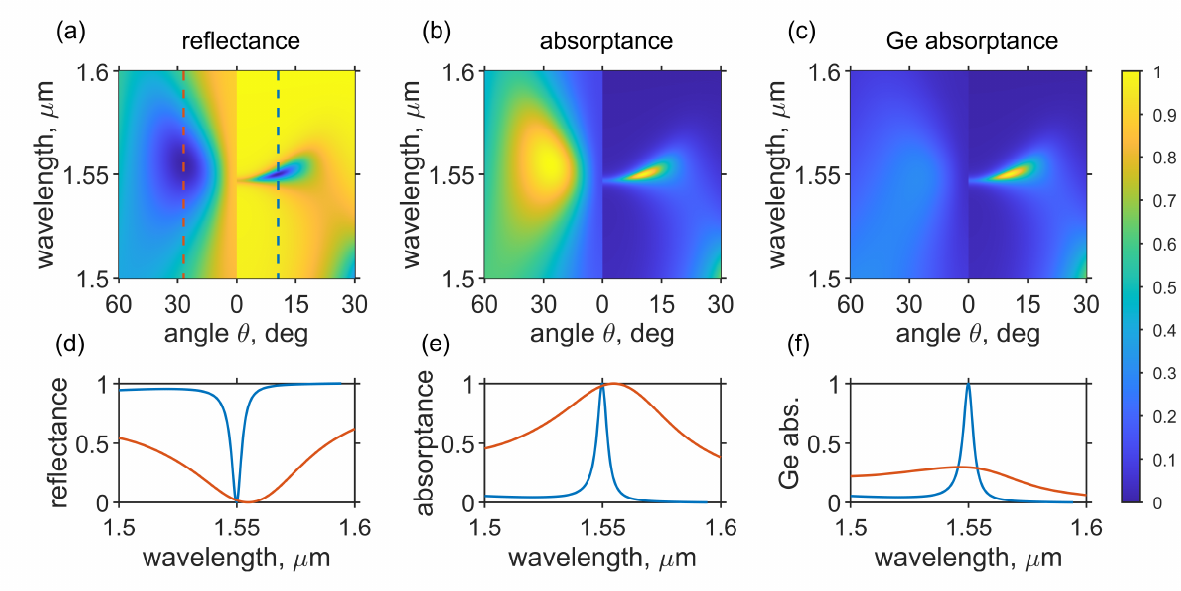}
\caption{(a) Reflectance, (b) absorptance spectra of the Au-based and PhC-based structures. (c) Absortance of the Ge in Au-based and PhC-based structures. For Au-based structure $H=93.5$~nm and $p=300$~nm. For PhC-based structure $H=200$~nm and $p=675$~nm. For both setups $L=350$~nm and $d=25$~nm. (d-e) Reflectance and absorptance spectra of the Au-based and PhC-based structures at 27 deg. and 10.5 deg., respectively. (f) Absorptance spectra of the Ge in both setups.}
\label{fig.Fig2}
\end{figure*}

It has been demonstrated in \cite{Saadabad21} that application of perfect mirror in the substrate of a BIC supporting structures 
makes it possible to set-up a perfect light absorber in the so-called critical coupling regime when the radiative and non-radiative $Q$-factors are equal to one another. In practice the mirror substrate for a dielectric structure can be implemented as an opaque metal film. All metals, however, exhibit
significant material losses so the radiation is absorbed not only
in the dielectric or semiconductor but in the mirror itself. Thus, the electromagnetic energy is wasted on heating the substrate rather than produce the desirable photoelectric effect.
In this letter we propose a set-up with a non-absorbing mirror which leads to $100 \%$ of incident radiation absorbed in the germanium (Ge) metasurface~\cite{Zhou2022}. In Table 1 of paper \cite{Odit2020}, the calculated total Q-factor ($1/Q_{total}=1/Q_{rad}+1/Q_{mat}$) of Ge is quite high at 1.5$\mu$m.
For designing high-Q metasurfaces at 1.5$\mu$m, it is more suitable than silicon in group IV. The system is depicted in Fig.~\ref{fig.Fig1}.
The structure consists of a periodic Ge grating placed on top 
of 1D photonic crystal. Later on the set-up will be referred to
as the PhC-based structure. For comparison we will consider a similar
Ge grating but placed on an Au substrate. The latter set-up
will be referred to as the Au-based structure and is also shown in Fig.~\ref{fig.Fig1}.

The Tamm metasurface structure has been proposed in our previous work~\cite{Bikbaev21} where we demonstrated that an off-$\Gamma$ BIC leads to emergence of critical coupling (CC) points in the parameter space of incident frequency $\omega$ and incident angle $\theta$. In~\cite{Bikbaev21} the critical coupling effect occurred as zero reflection due to tunnelling across the band-gap in the PhC substrate with a finite number of bilayers. Here we suppose that the substrate is thick enough to suppress the tunneling and expect that
the reflectance zeros are associated with the perfect light absorption. 

First, let us compare light absorption by the PhC- and Au-based structures. 
To carry out the comparison we defined the parameters of the structures such that both support an in-$\Gamma$ BIC at frequency $\lambda_{\rm{BIC}}=1.545~\mu$m. Corresponding parameters of the structure are presented in caption of Fig.\ref{fig.Fig1}.
The BICs do not couple with the incident light at the normal incidence. Therefore, to obtain the critical coupling one has to vary the angle of incidence. The increase of the angle of incidence results in a drop of the radiative $Q$-factor of the leaky band hosting the BIC until the radiation and material loss rates are equal to one another. This effect is visible in the reflectance spectra of the structures calculated by the finite difference time domain method shown in Fig.~\ref{fig.Fig2}~(a). In the PhC-based structure the
critical coupling is achieved at $\theta=10.5$~deg., thereas in 
the Au-based structure the critical coupling angle is $\theta=27$~deg. The Au-based structure is more lossy, hence the critical coupling is observed at a larger angle of incidence with a larger width of the resonance. 
One can see in Fig.~\ref{fig.Fig2}~(b) that in both structures $\approx100\%$ of incident radiation is absorbed at the critical coupling. However, as seen in Fig.~\ref{fig.Fig2}~(c) less than 30\% of energy is absorbed in Ge in the Au-based structure while in the PhC-based structure the absorptance 
in Ge is $A\approx 1$. 
Such a large difference in the absorption of the Ge is due to the fact that most of the radiation incident on the Au-based structure is absorbed in the plasmonic substrate, while in the PhC-based structure the substrate is all-dielectric. For the reader's convenience, the frequency dependencies of the absorption and reflection coefficients at the critical coupling are shown in Fig.~~\ref{fig.Fig2}~(d-f).

Notice that with an in-$\Gamma$ BIC the critical coupling is never obtained at the normal incidence. To engineer the perfect absorption in Ge at the normal incidence we return
to an off-$\Gamma$ BIC reported in~\cite{Bikbaev21}. 
Following~\cite{Bikbaev21} we describe the scattering spectrum of the Ge-based structure in the framework of the temporal coupled mode theory \cite{Fan03}
which yields the following solution for the reflection amplitude
\begin{equation}\label{reflection}
    r=\left(-1+\frac{2\gamma_1(\theta)}{i[\omega-
    \bar{\omega}(\theta)]+\gamma_1(\theta)+\gamma_2} \right),
\end{equation}
where $\omega$ is the frequency of the incident wave, $\bar{\omega}$ - the resonant frequency, $\theta$ - the angle of incidence, $\gamma_1$ is the loss rate due to coupling to radiation to the upper half-space, and $\gamma_2$ is the loss rate due to absorption. 
According to~\cite{Bulgakov17} the dispersion of $\gamma_1$ and $\bar{\omega}$ 
are given by
\begin{align}\label{Taylor}
    & \bar{\omega}=\omega_0-\alpha \theta^2 +{\cal O}(\theta^4), \nonumber \\
    & \gamma_1=\beta(\theta^2-\theta^2_{\rm BIC})^2 +{\cal O}(\theta^6),
\end{align}
where $\theta_{\rm BIC}$ is the angle corresponding to the off-$\Gamma$ BIC in the spectrum while $\alpha$ and $\beta$ are fitting parameters which can be obtained from numerical simulations. The absorption coefficient can be written
as
\begin{equation}\label{A}
A=\frac{4\gamma_1(\theta)\gamma_2}
{[\omega-\bar{\omega}(\theta)]^2+[\gamma_1(\theta)+\gamma_2]^2}
\end{equation}
\begin{figure}[h]
\centering\includegraphics[width=90mm]{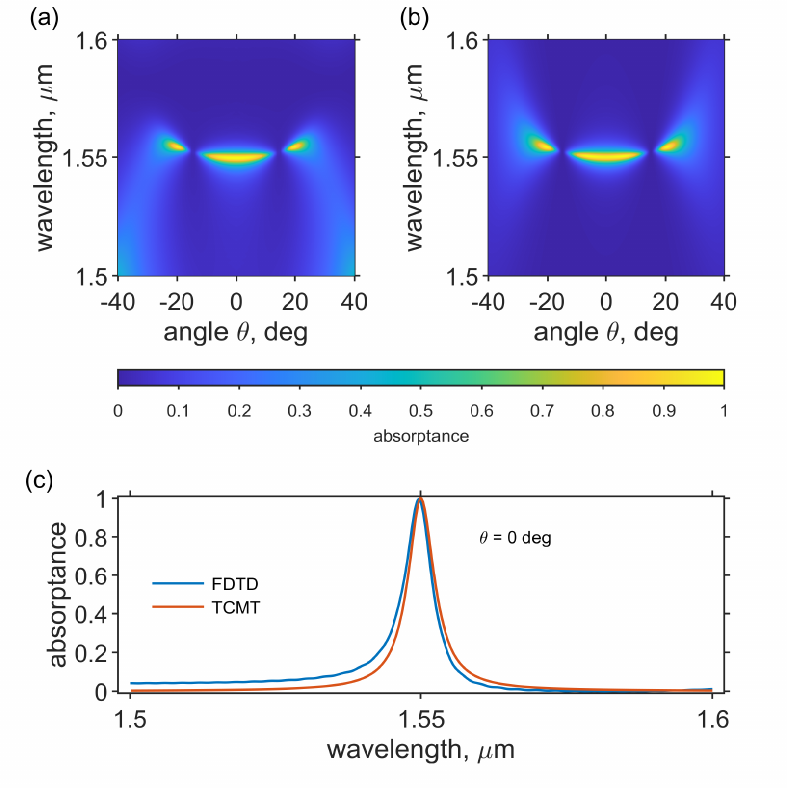}
\caption{Absorptance spectra of the PhC-based structure calculated by (a) FDTD and (b) TCMT theory. Comparison of the absorptance spectra of the structure calculated by two different approach at $\Gamma$-point. The period $p=620$~nm and $d=26.5$~nm. The other parameters are the same as for Fig.~\ref{fig.Fig2}. }
\label{fig.Fig3}
\end{figure}
The critical coupling points are topologically protected objects associated with phase singularities of the reflection amplitude~\cite{Bikbaev21}. For this reason, the effects of critical coupling and perfect light absorption
survive under variation of the system's parameters preserving all symmetries of the structure. Thus, by changing the parameters one can achieve the critical coupling in the $\Gamma$-point if
\begin{equation}
    \gamma_2=\beta\theta^4_{\rm{BIC}}.
\end{equation}
By varying the grating period we achieve
the in-$\Gamma$ critical coupling in the Ge-based structure
with period $p=620$~nm which led to the following values of
the TCMT parameters $\alpha\approx4.3\cdot 10^{-6}~\mu$m$^{-1}$deg$^{-2}$, $\beta\approx1.27\cdot 10^{-8}~\mu$m$^{-1}$deg$^{-4}$, $ \lambda_{\rm{BIC}}=1.55~\mu$m. In Fig.~\ref{fig.Fig3}~(a-b) we compare the numerical data against Eq.~(\ref{A}). The absorption spectrum at the normal incidence is shown in Fig.~\ref{fig.Fig3}~(c). One can see a good coincidence between the full-wave simulations and the TCMT approximation of the spectrum. 

In summary, we have demonstrated that application of a high-Q Tamm state makes in possible to engineer a perfect light absorber which provides
a $100\%$ absorption in a semiconductor grating. The proposed design
may be used at both normal and oblique incidence at the telecom wavelength. We believe that the reported wavelength and angle selectivity can be of use in small LiDAR detectors. 

This research was funded by the Russian Science Foundation (project no. 22-42-08003).
This work was supported by the Higher Education Sprout Project of the National Yang Ming Chiao Tung University and Ministry of Education and the National Science and Technology Council (NSTC 109-2628-E-007 -003 -MY3; 111-2923-E-007 -008 -MY3; 111-2628-E-007-021 ).

The authors declare no conflicts of interest.

\bibliography{sample,references,BSC_light_trapping}
\end{document}